\documentclass[apj]{emulateapj}
\pdfoutput=1
\usepackage[colorlinks,linkcolor={blue},citecolor={blue},urlcolor={red}]{hyperref}
\usepackage{epsfig} 
\usepackage{graphicx}
\usepackage{natbib} 
\usepackage{amsmath} 
\usepackage{amsfonts}
\usepackage{amssymb}

\newcommand{\hmpc}{\, h^{-1}\text{Mpc}}
\newcommand{\hsolar}{\, h^{-1} \text{M}_\odot}
\newcommand{\be}{\begin{equation}}
\newcommand{\ee}{\end{equation}}
\newcommand{\bc}{}
\newcommand{\bea}{\begin{eqnarray}}
\newcommand{\eea}{\end{eqnarray}}
\newcommand{\beast}{\begin{eqnarray*}}
\newcommand{\eeast}{\end{eqnarray*}}
\newcommand{\corr}[1]{\langle #1 \rangle}

\newcommand{\ctt}{\cos(2\theta_p)}
\newcommand{\coneone}{c_{11}}

\begin{document}

\title{Halo intrinsic alignment: dependence on mass, formation time and environment}

\author{Qianli Xia\altaffilmark{1,2}, Xi Kang\altaffilmark{1},
  Peng Wang\altaffilmark{1,2}, Yu Luo\altaffilmark{1}, Xiaohu Yang\altaffilmark{3}, Yipeng Jing\altaffilmark{3}, Huiyuan Wang\altaffilmark{4}, Houjun Mo\altaffilmark{5,6}}

\altaffiltext{1}{Purple Mountain Observatory, the Partner Group of MPI f$\ddot{\text{u}}$r Astronomie, 2 West Beijing Road, Nanjing 210008, China}
\altaffiltext{2}{Graduate School, University of the Chinese Academy of Science, 19A, Yuquan Road, Beijing 100049, China}
\altaffiltext{3}{Center for Astronomy and Astrophysics, Shanghai Jiao Tong University, Shanghai 200240, China}
\altaffiltext{4}{Key Laboratory for Research in Galaxies and Cosmology, Department of Astronomy, University of Science and Technology of China, Hefei, Anhui 230026, China}
\altaffiltext{5}{Astronomy Department and Center for Astrophysics, Tsinghua University, Beijing 10084, China}
\altaffiltext{6}{Department of Astronomy, University of Massachusetts, Amherst MA 01003305, USA}

\email{kangxi@pmo.ac.cn}
\begin{abstract}

In this paper we use high-resolution cosmological simulations to study halo
intrinsic  alignment and its  dependence on mass,  formation time and large-scale environment.  In agreement with previous studies using N-body
simulations, it  is  found that  massive  halos have  stronger alignment. For
the first time, we find that for given halo mass, older halos have stronger
alignment and halos in cluster regions also have stronger alignment than those
in filament.  To model these dependencies we extend the  linear alignment
model with inclusion of halo bias and find  that the halo alignment with its
mass and formation time dependence can  be explained by halo bias. However,
the model can not  account for the environment dependence, as it is found that
halo bias is lower in clusters and higher in filaments. Our results suggest that
halo bias and environment are independent factors in determining halo
alignment. We also study the halo alignment correlation  function and find
that  halos are strongly clustered along  their major axes  and less clustered
along  the minor axes. The  correlated halo alignment can  extend to scale as
large as $100h^{-1}$Mpc  where  its feature  is  mainly  driven by  the
baryon acoustic oscillation effect.

\end{abstract}
\keywords{dark matter --- large-scale structure of universe ---
  galaxies: halos --- galaxies: formation --- methods:  statistical}

\section{Introduction} 
\label{sec:introduction}

Observational  data from  large sky  surveys have  clearly  shown that
galaxies are aligned with each  other and also with the matter distribution
on large-scale  structure. On galactic scales,  the satellite galaxies
are  aligned  with  the  major  axis  of  the  central  galaxy  (e.g.,
\cite{2004MNRAS.348.1236S},                 \cite{2005ApJ...628L.101B},
\cite{2005MNRAS.362..711Y}). On  scales larger  than  a few
$\hmpc$,  the  spin of  spiral  galaxies  are  also correlated  (e.g.,
\cite{2000ApJ...543L.107P}; \cite{2011ApJ...732...99L}), so is for the
shape of galaxies (\cite{2002MNRAS.333..501B},
\cite{2004MNRAS.353..529H},                 \cite{2004MNRAS.347..895H},
\cite{2009ApJ...694..214O},                 \cite{2009RAA.....9...41F},
\cite{2011A&A...527A..26J},                 \cite{2013MNRAS.431..477J},
\cite{2015MNRAS.450.2195S}).  The  galaxy shape  alignment can  be extended  to very
large   scale    around   70-100$\hmpc$   (\cite{2012MNRAS.423..856S},
\cite{2013ApJ...770L..12L}). Galaxies  are also found to align with the  cosmic web (e.g.,
\cite{2007ApJ...671.1248L},         \cite{2013ApJ...779..160Z}), though  with
dependence       on       galaxy       mass       and       morphology
(\cite{2013ApJ...775L..42T}). For  a  recent review on the various kinds  of galaxy alignment, we refer the readers to the papers by \cite{2015SSRv..193....1J} and \cite{2017SSPMA..47d9803K} (but in chinese language).

Among the above various galaxy alignments, the shape correlation between galaxies on large scales is of great importance as it is a major contamination to weak lensing surveys and precision cosmology requires a good understanding of galaxy alignment (e.g., \cite{2012MNRAS.424.1647K}). N-body simulations have been extensively used to study the intrinsic alignment (IA) of dark matter halos since the beginning of this century (e.g., \cite{2000ApJ...545..561C}; \cite{2000MNRAS.319..649H}; \cite{2002MNRAS.335L..89J}). These studies found that halo IA is strong in massive halos and a useful fitting formula is given in \cite{2002MNRAS.335L..89J}. To explain the observed galaxy IA at low-redshifts, a mis-alignment between galaxy shape and halo has to be included (e.g., \cite{2007MNRAS.378.1531K}; \cite{2009RAA.....9...41F}; \cite{2009ApJ...694..214O}; \cite{2013ApJ...770L..12L}; \cite{2013MNRAS.431..477J}). However, it is not clear how the mis-alignment between halo and galaxy shape should depend on galaxy properties. In recent years, cosmological hydro-dynamical simulations are also used to directly measure galaxy IA and its dependence on galaxy mass, luminosity and redshift (e.g., \cite{2014MNRAS.444.1453D}; \cite{2015MNRAS.448.3522T}; \cite{2015MNRAS.454.3328V}; \cite{2016MNRAS.461.2702C}; \cite{2017MNRAS.468..790H}). These studies have found in common that galaxy IA depends on luminosity and morphology. However, due to the difference of the adopted cosmology and the uncertainty of implemented star formation physics, the predicted galaxy IA are still different in detail from the hydro-dynamical simulations. For a review on the progress of measuring galaxy/halo IA using simulations, please refer to the paper by \cite{2015SSRv..193...67K}.

For a better understanding of the physical origin of halo/galaxy IA, we need a theoretical model which can recover the halo/galaxy IA as seen in simulations. The linear alignment model (\cite{2004PhRvD..70f3526H}) is developed based on the tidal theory (e.g., \cite{2001MNRAS.320L...7C}) with major assumption that halo/galaxy shape is determined by the local tidal field. Subsequently, the linear model is improved on non-linear scales (\cite{2011JCAP...05..010B}) and a more comprehensive halo model is also developed (\cite{2010MNRAS.402.2127S}). However, in these analytical models the halo bias is neglected, and galaxy bias is presented only in the second order of power spectrum for galaxy IA. Thus, it is still not clear if the analytical model can  account for the simulated galaxy IA in detail, such as the dependence on galaxy formation time and environment. As a further step to model the galaxy IA in detail, in this paper we use N-body simulation to study halo IA, especially we present the first measurement of halo IA with dependence on halo formation time and cosmic environment. We also investigate if the linear model with inclusion of halo bias can explain the halo IA with dependence on formation time and cosmic environment.

Our paper is organized as follows. In  Sec.\ref{sec:method}  we
introduce our  simulation data, the  method to clarify the  large scale
structure  and the modified analytical  model  to describe  the  halo IA.   In
Sec.\ref{sec:c11} we  present the simulation  results on halo  IA with
dependence on  halo mass, formation time  and large-scale environment,
and we also show how our  slightly modified linear tidal model can fit
the mass and formation time  dependence of halo  IA. In Sec.\ref{sec:ACF} we  present the results on  the halo  alignment correlation functions  and investigate
the  origin behind the oscillation  of halo  IA on  very large  scales.  In
Sec.\ref{sec:Conclusion} we summarize  our results and briefly discuss
their implications.

\section{Methodology}
\label{sec:method}
In this section, we introduce the simulation data, calculation of halo IA and definition of the large-scale environment of the halo. We will also show the slightly modified linear tidal model to describe the halo IA in 3D coordinate measured from our simulations.

We use two cosmological N-body simulations in this paper. One is the Pangu simulation, carried out by the Computational Cosmology Consortium of China (\cite{2012ApJ...761..151L}, hereafter this simulation is abbreviated as C4), which simulated the evolution of the universe in a cubic box with each side of 1000$\hmpc$. The other simulation is a part of the ELUCID project (\cite{2014ApJ...794...94W,2016ApJ...831..164W}, \cite{2016RAA....16..130L}) with box-size 500$\hmpc$ (Hereafter L500). Both simulations are run by the GADGET-2 code (\cite{2005MNRAS.364.1105S}) with $3072^3$ particles and the cosmological parameters can be found in Table \ref{table1}. Note that the cosmological parameters in the two simulations are very close and we do not expect significant difference on our results, and we will also show comparison between them.

\begin{table}[t]
  \caption{Simulation Parameters}
  \begin{tabular}{ c | c  c  c  c  c  c}
    \hline
    Name & $\Omega_M$ & $\Omega_\Lambda$ & $\Omega_b$ & h & L ($h^{-1}$Mpc) & $m_p (\hsolar)$ \\ \hline
    L500 & 0.28 & 0.72 & 0.045 & 0.7 & 500 & $3.4\times 10^{8}$ \\ \hline
    C4 & 0.26 & 0.74 & 0.045 & 0.7 & 1000 & $2.5\times 10^{9}$ \\ \hline
  \end{tabular}
  \label{table1}
\end{table}

Using the friends-of-friends (FOF) algorithm with a linking length $b=0.2$ of the mean particle separation, we identified dark matter halos in both simulations. For each halo, we project its dark matter particles along one axis of the simulation box and calculate the halo's reduced 2D moment of inertia tensor $I_{ij}$ (\cite{2005ApJ...627..647B}) as,
\be
I_{ij} = \sum_k \frac{m_\alpha x_{k,i} x_{k,j}}{x^2_k}, \ \text{ with}  i,j \in \{1,2\} ,
\ee
where $x_k$ is the distance of the $k$th particle to the halo center which is set as the position of the most bound particle. The eigenvalues $\lambda_1,\lambda_2~(\lambda_1>\lambda_2)$ of the inertia tensor then define the axis ratio $q=\lambda_2/\lambda_1$. Then, following
\cite{2000MNRAS.319..649H} and \cite{2002MNRAS.335L..89J}, we define the halo shape by the ellipticity vector,
\be
{\boldsymbol\epsilon} = \frac{1-q^2}{1+q^2}(\cos2\alpha, \sin2\alpha)^T = (\epsilon_1, \epsilon_2)^T ,
\ee
which in the complex form reads, 
\be
\epsilon = \epsilon_1 + {\rm i} \epsilon_2 = \frac{1-q^2}{1+q^2} \exp\left({\rm i}2\alpha\right) .
\label{eq:complex-shear}
\ee
where $\alpha$ is the position angle of the halo measured anti-clockwise from x-axis.

As pointed by earlier studies (\cite{2002MNRAS.335L..89J}; \cite{2005ApJ...627..647B}), a lower limit of particle number ($\approx$ 300 particles) in the halo has to be used to get a converged shape measurement. To ensure a safer measurement, in this work we select halos with more than 500 particles.  In addition, it is found that the halo shape is also dependent on the linking length to find halo in the FOF algorithm. \cite{2000ApJ...545..561C} had shown that this can lead to the discrepancy of halo IA at the low-mass end. We will later show comparison of our results with \cite{2002MNRAS.335L..89J}.

To define the local large-scale environment of a halo,  we use the Hessian matrix of the density field at the position of the halo (e.g., \cite{2007A&A...474..315A}). Adopting the Cloud-in-Cell (CIC) algorithm, the density field is smoothed via a length of $2\hmpc$ and the Hessian matrix of the smoothed density field $\rho_s$ at halo position is described as,
\begin{align}
  H_{ij} = \frac{\partial \rho_s}{\partial x_i \partial x_j} ,    
\end{align}
The eigenvalues of the Hessian matrix define the large-scale environment of the halo, and it can be classified into cluster, filament, sheet and void, depending on the sign of the eigenvalues. By ordering the eigenvalue as $\lambda_1 > \lambda_2 > \lambda_3$, the environment is linked as,
\beast
(-,-,-) &\rightarrow& \text{cluster}\\
(+,-,-) &\rightarrow& \text{filament}\\
(+,+,-) &\rightarrow& \text{sheet}\\
(+,+,+) &\rightarrow& \text{void}
\eeast

\begin{figure}[t]
  \includegraphics[width=\columnwidth]{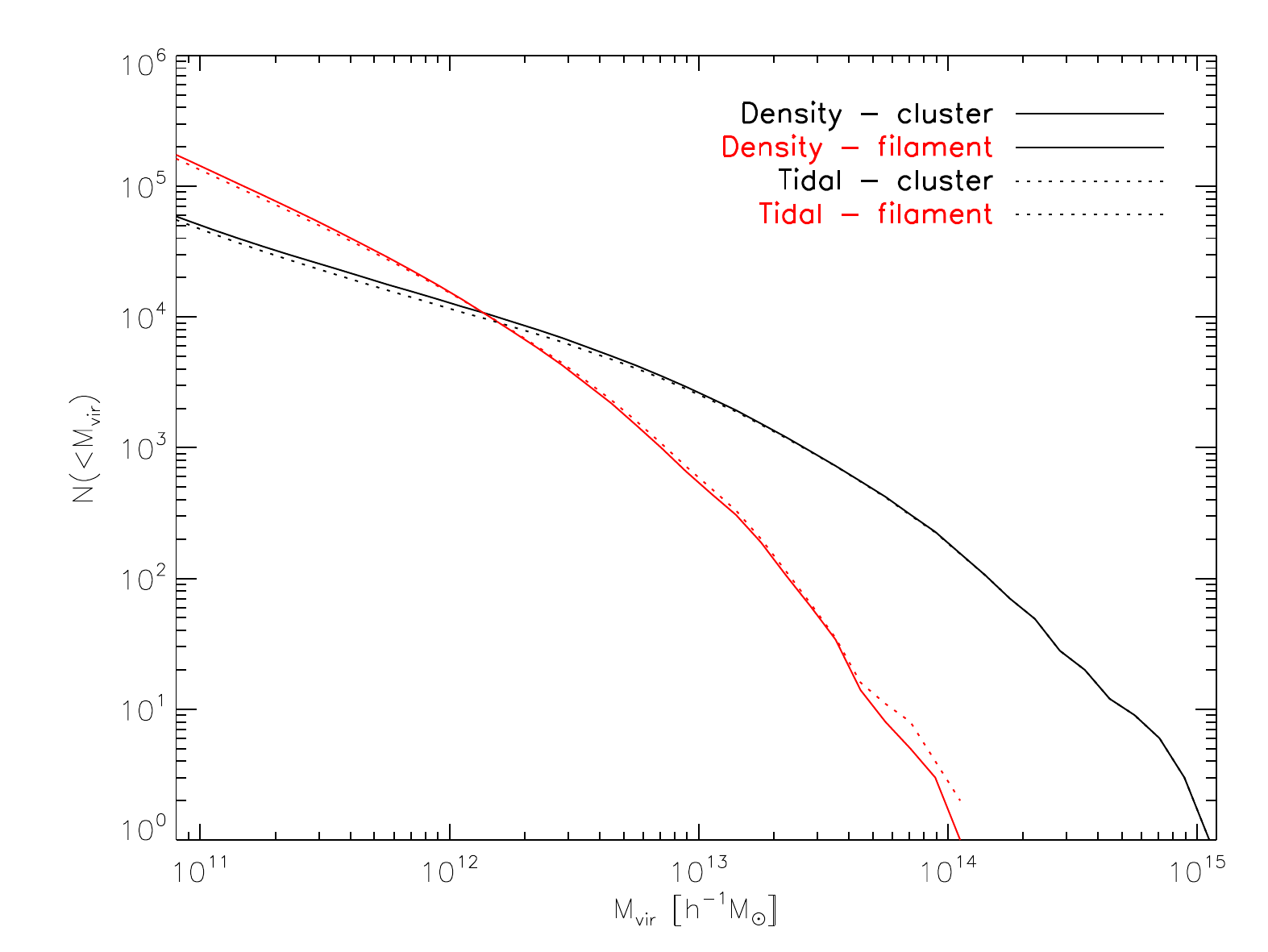}
    \caption{The mass function of halos in clusters (black) and filaments (red)
      from L500 using Density-Hessian classification (solid) and Tidal-Tensor classification (dotted).}
    \label{fig:density_TTT}
\end{figure}

We note that this way of defining LSS environment is different from the tidal environment defined in \cite{2007MNRAS.375..489H} in which the Hessian of the potential (i.e., tidal field) is used and classification is done using an opposite signature of eigenvalues. In Fig.~\ref{fig:density_TTT}, we plot the halo mass function in cluster and filament environment from the two different classification schemes, and in Table \ref{table2}, we list the ratio of each LSS environment under the two classification methods. It is seen from Fig.~\ref{fig:density_TTT} that the halo mass functions agree well in the two environments. By checking the environment of each halo, we find that 77\% of all halos keep their environment unchanged and 23\% of them changed when changing the classification method. For this paper, we will adopt the density-Hessian classification scheme. We will still be using the density Hessian as ``tidal field'' in our context, but the reader should be aware of the mentioned difference when comparing with previous studies which adopts tidal tensor. We also direct the reader to \cite{2017arXiv170503021L} for detailed comparison and discussions on different LSS classification schemes.

\begin{table}[h]
  \caption{LSS Environment Ratios (\%)}
  \begin{tabular}{ c | c  c  c  c }
    \hline
    ~~              & Cluster & Filament  & Sheet   & Void   \\
    \hline
    Density Hessian & 20.0359 & 61.8152   & 17.7563 & 0.3926 \\
    \hline
    Tidal Field     & 19.3583 & 58.0268   & 21.1922 & 1.4227 \\
    \hline
    In Total        & Same    & Different & ~~      &~~      \\
    \hline
                    & 76.9173 & 23.082    & ~~      &~~      \\
    \hline
  \end{tabular}
  \label{table2}
\end{table}

Basically the environment of the halo is related to its mass: usually massive halos live in cluster regions and low-mass halos live in sheets and voids. The halo shape is also closely related to the eigenvector of the environment, for a short summary of how halo shape is related to the environment and its physical origins, we refer the reader to the paper by \cite{2015ApJ...813....6K} and \cite{2017MNRAS.468L.123W} and references therein.

Now we describe how the halo IA is usually defined in the weak lensing theory. Following the conventional description (\cite{2000MNRAS.319..649H}; \cite{2002MNRAS.335L..89J}), for a pair of halos (halo 1 and halo 2), let $w_1 = x_1 + {\rm i}y_1$ and $w_2 = x_2 + {\rm i}y_2$ 
be projected positions of two halos in the complex plane, and call the direction defined by 
$\varphi = \arg(w_2-w_1)$ 
(anti-clockwise from the positive real axis) as the separation direction . Then, for this pair of halos we can define the tangential shear $\gamma_t$ and the cross shear $\gamma_x$
\be
\gamma_t = -\operatorname{Re}\left( \epsilon \exp(-2{\rm i}\varphi)\right),~~ \gamma_\times = -\operatorname{Im}\left( \epsilon \exp(-2{\rm i}\varphi)\right) ,
\label{eq:def_shear}
\ee
where $\epsilon$ is the complex shear of a given halo defined in Eq.~\ref{eq:complex-shear}. The halo IA is usually defined as $c_{11}(r) = \langle\gamma_{t}(0)\gamma_{t}(r)\rangle$, sometimes also labeled as $\eta_1(r)$, and $\langle\gamma_{\times}(0)\gamma_{\times}(r)\rangle$ is labeled as $c_{22}(r)$. 

By assuming that galaxy shapes are determined by their local tidal shear, the linear alignment model
(
\cite{2004PhRvD..70f3526H})
suggests that galaxy ellipticity is: 
\begin{align}
\gamma^I_{(t,\times)} = -\frac{C_1}{4\pi G}(\nabla_x^2 - \nabla_y^2,
2\nabla_x\nabla_y)S[\Psi_P] ,
\label{eqn:Hirata}    
\end{align}

Where $C_{1}$ is a free parameter which can be determined by fitting the observed galaxy IA. Up to now, most studies have determined the value of $C_1$ by fitting the IA correlation of low-z massive galaxies (luminous red galaxies) as a whole. So it is not clear how the normalization will vary as a function as galaxy luminosity. Also as the fitting is for galaxy IA, it is not clear if the tidal model (Eq.~\ref{eqn:Hirata}) can describe the IA of dark matter halo and its mass dependence.

To investigate the impact of galaxy IA  on probing cosmology, it is important
to measure the  IA signal in the shear correlation  function, which is
defined as,
\begin{align}
\xi_\pm(\theta) = \corr{\gamma_t\gamma_t}(\theta) \pm \corr{\gamma_\times\gamma_\times}(\theta)    
\end{align}
where $\theta$ is the angular separation on the sky. \cite{2011JCAP...05..010B} (Eq.~3.5 therein) showed that under the linear alignment model, the projected tangential and cross components of the shear correlation function to the linear order is,
\begin{align}
    \corr{\gamma\gamma}^{LA}_{(tt,\times\times)}(r_p) &= \frac{1}{{2\pi^2}} \left(\frac{C_1 \rho_{m,0}}{D(z)}\right)^2 \int d\kappa dk_z \frac{\kappa^5}{k^4 k_z}P_{\delta}(k,z) \nonumber \\ 
    & \times \sin(k_z \Pi_{\rm max})\left[J_0(\kappa r_p) \pm J_4(\kappa r_p)\right].
\end{align}
However, for $\coneone$ in \cite{2002MNRAS.335L..89J}, the correlation of projected shapes are calculated at each 3D distance between halo pairs. We explicitly calculate our model prediction for $\coneone$ by first performing the integral $\xi_\pm(r) = \int_0^\pi\xi_\pm(r_p = r\sin\phi, \Pi=r\cos\phi)~{\rm d}\phi$. Then, after analytically integrate over azimuthal angles, we obtain the final formula

\begin{align}
\xi_{-}(r) = \left( \frac{C_1 \rho_{m,0}}{D(z)} \right)^2 \frac{1}{4\pi}\int_0^\infty {\rm d}k~\frac{\mathcal{P}_\delta(k)}{k}~\sqrt{\frac{2\pi}{kr}}~\frac{3\pi}{8}~J_{\frac{9}{2}}(kr) ,
\end{align}
and, 
\begin{align}
\xi_{+}(r) & = \left( \frac{C_1 \rho_{m,0}}{D(z)} \right)^2 \frac{1}{4\pi} \int_0^\infty {\rm d}k~\frac{\mathcal{P}_\delta(k)}{k}~\sqrt{\frac{2\pi^3}{kr}} \nonumber \\
& \times \left(\frac{9}{280}J_{\frac{9}{2}}(kr)  + \frac{4}{21}J_{\frac{5}{2}}(kr) + \frac{56}{105} J_{\frac{1}{2}}(kr)\right) ,
\end{align}

where $J_{n+\frac{1}{2}}$ is the Bessel function of half-integer order. Hence, by relation $\coneone(r) = \corr{\epsilon^h_+\epsilon^h_+}(r)= \frac{1}{2}(\xi_+(r) + \xi_-(r))$, we now have the prediction of $\coneone(r)$ from the linear alignment theory.

Since we are calculating the shear correlation functions at halo positions, we will be using $\mathcal{P}_{\delta,h}$ instead of $\mathcal{P}_\delta$ in the above equations. For halos at fixed mass, the power spectrum at halo position is $\mathcal{P}_{\delta,h}(M) = b_h^2(M)~\mathcal{P}_\delta$, where $b_h$ is the halo bias.
The fact that we will be calculating ellipticity correlation of halos within different mass ranges suggests that we should extend to, $\mathcal{P}_{\delta,h}(M_h\ge M)$, which by conditioning on halos with mass $M_h\ge M$ and to linear order: 
\begin{align}
\mathcal{P}_{\delta,h}(M_h\ge M) = b_h^2(M_h\geq M)\mathcal{P}_\delta ,    
\end{align}

Therefore, we get the mass-dependent $\coneone$ for halos
\begin{align}
\coneone(r,\ge M) = \left(\frac{ \int_M^\infty b_h(M') \Phi(M') {\rm d}M' }{ \int_M^\infty \Phi(M') {\rm d}M' } \right)^2 \corr{\epsilon^h_+\epsilon^h_+}(r),
\label{eqn:theory}    
\end{align}
where $\Phi(M)$ is the halo mass function which can be obtained for given cosmology.

\section{Dependencies of Halo Intrinsic Alignments}
\label{sec:c11}
\subsection{Mass dependence of halo IA}
\label{sec:massdep}
In  Sec.\ref{sec:method} we  have outlined  the theoretical  model for
halo  IA and in this section, we  compare the simulation results to model
predictions to see if the linear model can fit the mass dependence of halo IA.

\begin{figure}[t]
  \includegraphics[width=\columnwidth]{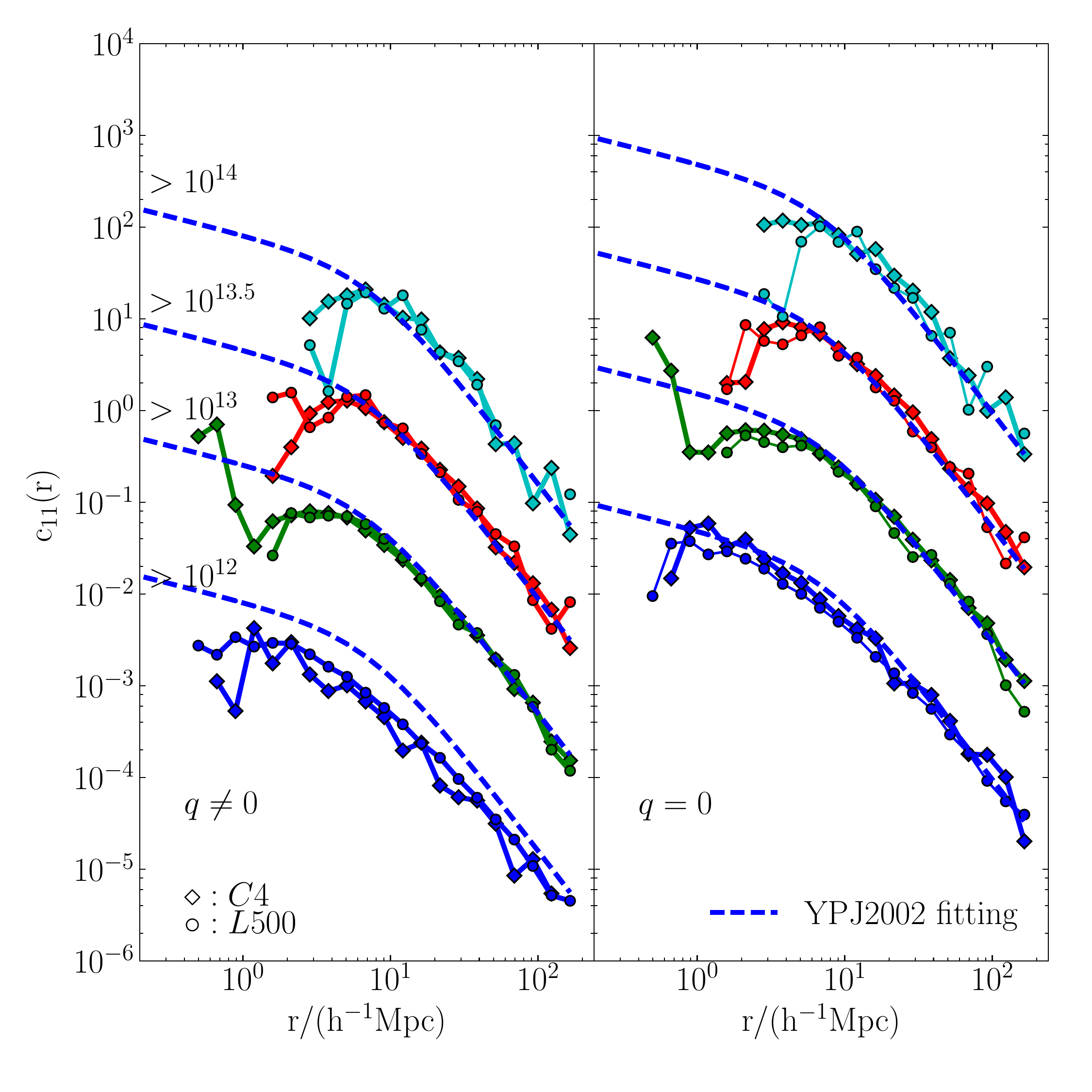}
    \caption{The ellipticity correlation  function $\coneone$ of halos
      from L500 (circles connected  with solid lines) and C4 (diamonds
      connected with solid lines) simulations at $z=0$. The results for
      different halo mass bins ($\log(M/(h^{-1}M_\odot))>12, 13, 13.5,
      14$ (from bottom  to top) have been multiplied by  1, 10, 100 and
      1000 respectively for clarity.  The fitting formula from Jing (2002) 
      is shown as  blue dashed lines in both  panels. Left: the ellipticity weighted correlation. Right:
      no ellipticity included, only halo orientation is correlated.}
    \label{fig:$\coneone$_YPJ}
\end{figure}

First  we note  that, as  pointed by  \cite{2002MNRAS.335L..89J}, when
calculating  $\coneone$  from   Eq.~\ref{eq:def_shear},  not  only  the
direction of the halo shape is correlated, the ellipticity of the halo
is also taken into account (the term $(1-q^2)/(1+q^2)$), so is for the
theoretical description by Eq.~\ref{eqn:Hirata}.  So basically that is the
ellipticity weighted $\coneone$. In fact we can also only consider the
orientation  correlation   by  assuming  that   $q=0$  (unweighted  by
ellipticity)  in the  simulations. In  Fig.\ref{fig:$\coneone$_YPJ} we
show the measured weighted and unweighted $\coneone$ for halos larger
than given mass in the left and right panel. The circles are from L500
simulation  and diamonds  are from  C4.  For clarity,  the result  for
different halo mass  bin has been shifted by  different factors. It is
found that the  results from the two simulations  agree quite well for
both weighted  and unweighted correlations. For the  following we will
only show results from the L500 simulation unless otherwise stated.

\cite{2002MNRAS.335L..89J} measured the weighted $\coneone$ from their
simulation, and found  it can be well fitted by  a simple formula with
mass dependence as,
\begin{align}
\label{eqn:YPJ2002}
c^\text{YPJ}_{11}(\geq M_\text{h}; r) = \frac{3.6\times10^{-2}\left( \frac{M_\text{h}}{10^{10}\hsolar}\right)^{0.5}}{r^{0.4}\left( 7.5^{1.7} + r^{1.7} \right)} ,    
\end{align}
We     plot      their     fitting     as      dashed     lines     in
Fig.~\ref{fig:$\coneone$_YPJ}. It  is worth noting  that their fitting
formula  was  intended  for  FoF  halos with  a  linking  length  with
$b=0.1$. As pointed in Jing (2002)  the mass of FOF halo using $b=0.1$
is about half of the FOF halo mass with $b=0.2$. Therefore, the dashed
lines here are plotted using the  fitting formula but with half of the
halo  mass in our  simulation (with  $b=0.2$).  It  is found  that the
fitting formula can  also well fit our results except  at the low mass
end    where    our    $\coneone$    is    lower    than    that    of
$\coneone^\text{YPJ}$.   It  indicates   that  $\coneone$  in  our
simulation  has a  slightly  stronger mass  dependence.   As shown  by
\cite{2000ApJ...545..561C},  such a slight  difference is  expected as
the  measured halo  shape is  slightly different  with  different link
length,  especially  for  low-mass  halos.  In  addition,  given  the
additional  difference in cosmological  parameter, simulation  box and
resolution, we believe such a slight difference is acceptable. The right
panel  shows  that  without  considering  the  halo  ellipticity,  our
$\coneone$  agrees with  the fitting  formula much better.   We also note that for the unweighted $\coneone$ the fitting formula is multiplied by a factor of $\sim 6$ to produce good match with the data, similar to the finding of \cite{2009ApJ...694..214O}. The fact that the fitting formula at $q=0$ works well also confirms that the difference seen in the left panel for low mass halos mainly comes from randomness of halo ellipticities at low mass.

\begin{figure*}
 \begin{center}
   \epsfig{figure=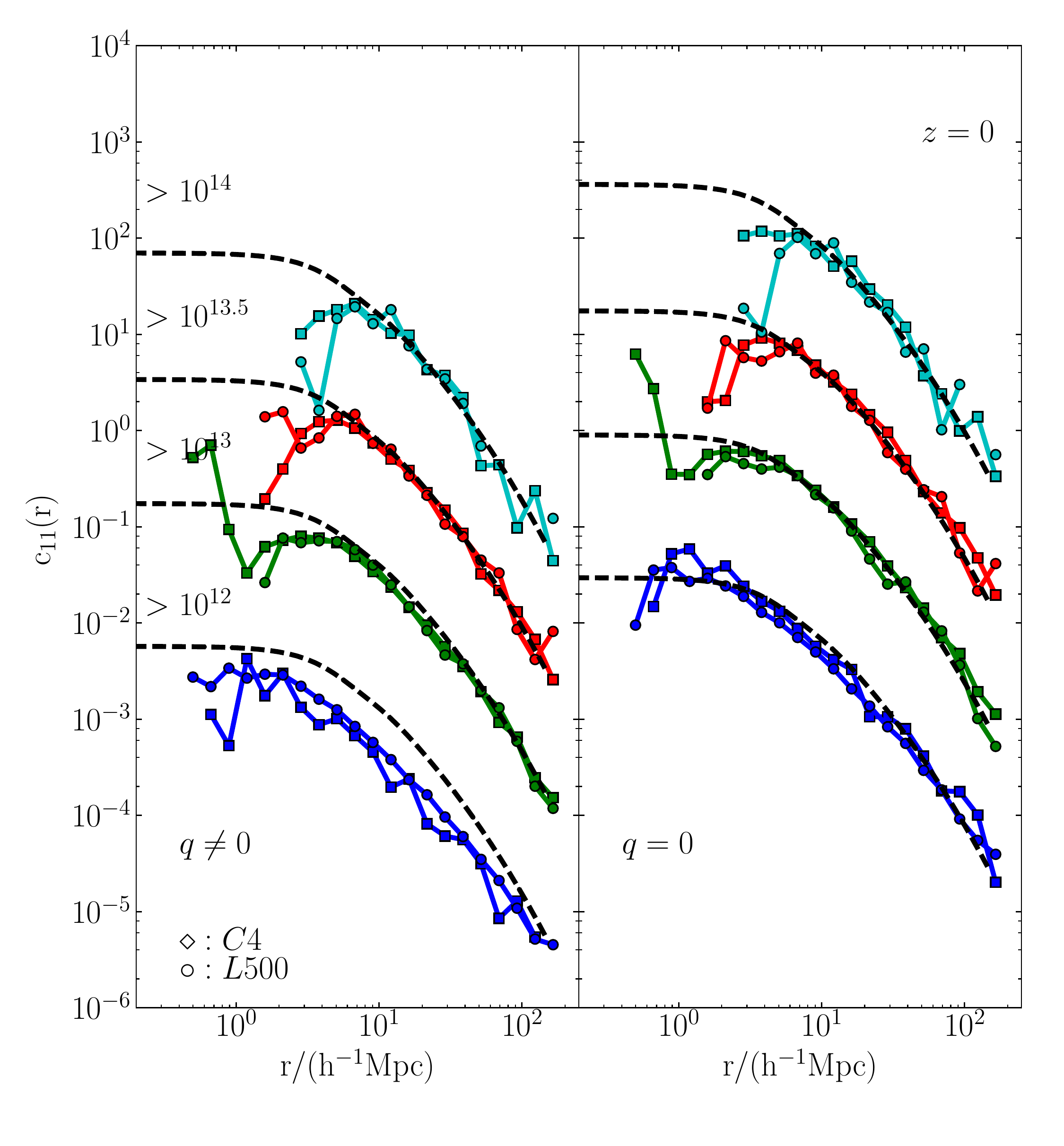,width=0.47\hsize}
   \epsfig{figure=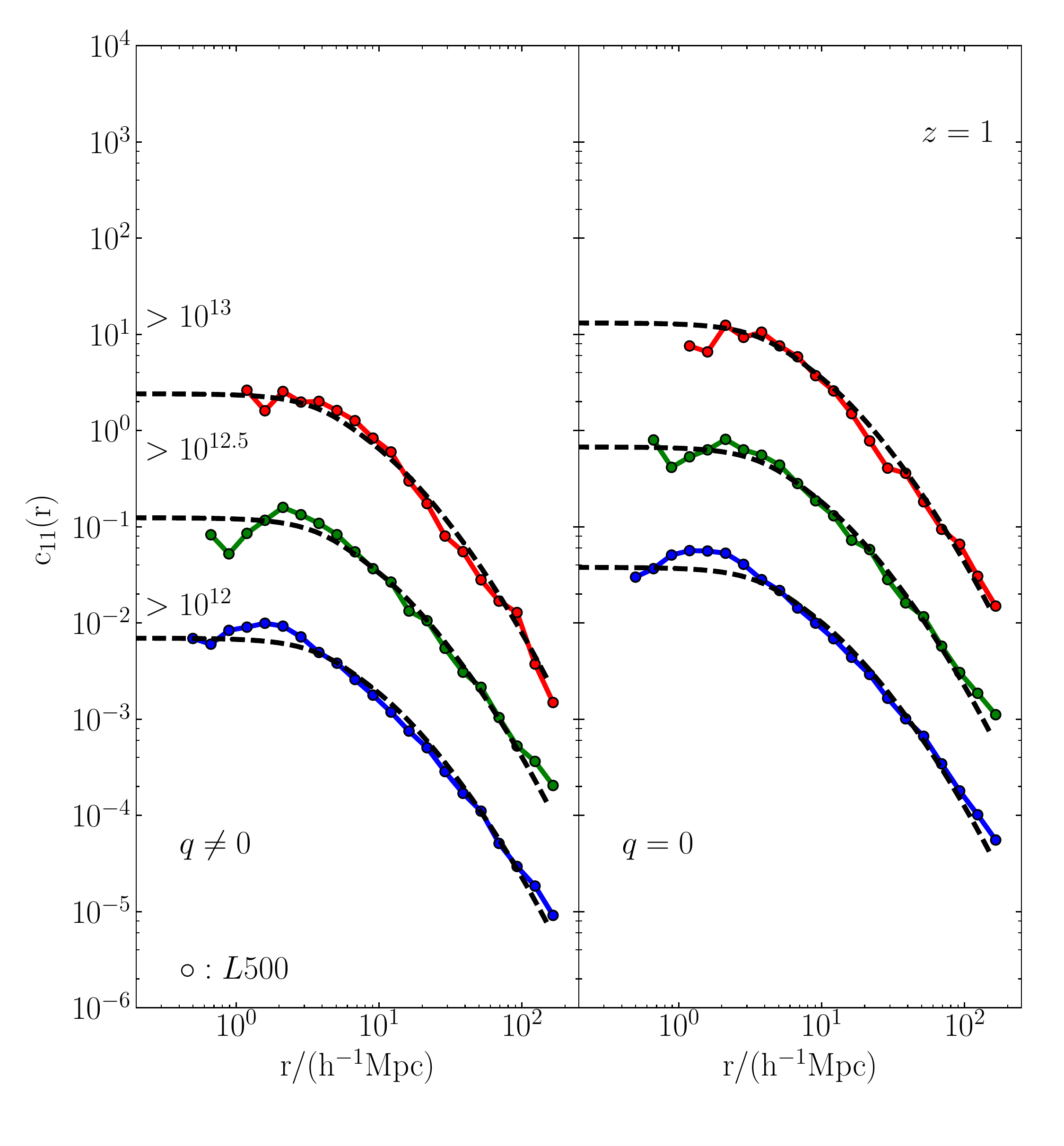,width=0.47\hsize}
 \end{center}
 \caption{The ellipticity correlation function $\coneone$ of halos at two redshifts. Left: correlation at $z=0$. Similar to Fig.~\ref{fig:$\coneone$_YPJ}, but the dashed lines are the theoretical model predictions. Right panel: the correlation at $z=1$.}
 \label{fig:bias_model}
\end{figure*}

In Fig.~\ref{fig:bias_model}  we compare our simulation  results to the
theoretical  model.   The  left   panel  is  the   same  as   that  of
Fig.~\ref{fig:$\coneone$_YPJ}, but now the  dashed lines are the model
predictions  (Eq.~\ref{eqn:theory}).  For   the  theoretical  model  we
adopted  the halo  bias from  \cite{2014MNRAS.442.1930P} and  the halo
mass function  $\Phi(M)$ is taken  from our simulation. For  the power
spectrum   we  used   the  nonlinear   power  spectrum   from  halofit
\cite{2000MNRAS.318.1144P}     adhering    to   the  L500    cosmological
parameters. In  addition, to better  match the real power  spectrum in
simulation     we      added     Cloud-In-Cell     window     function
\cite{2008ApJ...687..738C}  with a  2$\hmpc$ smoothing  kernel  in the
computation. We tune the  free model parameters for weighted and unweighted $\coneone$ respectively to
better match  the simulation results for the highest mass bin  (upper line). It is  seen that the
simulation results can be well matched by the theoretical predictions,
indicating that  it is  mainly the halo  bias accounting for  the mass
dependence  of  halo IA,  $\coneone$.  Again  we  find that  the  mass
dependence  in the  simulation is  slightly stronger  than  the linear
tidal  model.  The  right  panel  shows  a  better  agreement  on  the
unweighted  $\coneone$ with  the model.  The better  agreement  on the
unweighted $\coneone$ indicates that the correlation between direction
of the halo  (major axis) is better described  than the ellipticity by
the linear tidal model.

In the right panel of Fig.~\ref{fig:bias_model} we plot the $\coneone$
at $z=1$ as  an additional test of the  linear tidal model. Amazingly,
the agreement at higher redshift is also good and even better than the
$z=0$ results.  This is also  expected as it  is known that  the tidal
field  at  earlier  time  can   be  better  described  by  the  linear
theory.  Following equation  (B.5)  in \cite{2011A&A...527A..26J},  we
also note that the ratio  between the free parameter in our unweighted
model   $\frac{C_1(z=1)  \rho_{m,0}}{D(z=1)}/\frac{C_1(z=0)  \rho_{m,0}}{D(z=0)}
\approx 1.40$ is  slightly lower than the ratio from the linear growth factor
${D(z=0)}/{D(z=1)}  \approx 1.61$.  This indicates  that  the redshift
evolution  can be roughly captured  by the  power spectrum  at different
redshift, yet  there is still other  factors which is  absorbed in the
normalization factor $C_1$ which should slightly depend on redshift.

\subsection{Formation time and environment dependence}

It is well known that  halo properties, such as concentration and
bias,   are   closely   related   to halo  formation   time   (e.g.,
\cite{1997ApJ...490..493N};   \cite{2004MNRAS.355..819G}).    In  this
section we investigate  whether the halo IA is  also dependent on halo
formation time. Here,  we use the most common  definition of formation
time, $z_f$, at  which the mass assembled in the  main progenitor of a
halo  is  half  of  its   present  (z=0)  mass,  i.e.,  $M_h(z=z_f)  =
\frac{1}{2}M_h(z=0)$.  As  there is a strong  correlation between halo
formation  time and  its mass  (e.g.,  \cite{1997ApJ...490..493N}), we
select halos in a few narrow  mass bins and divide the 20\% oldest and
youngest halos into the old and young  samples. In addition, as
we will soon see that halo  IA is also dependent on their environment,
we refine  the old and young halo  samples to make sure  they have the
same   distributions   of   cosmic   environment  as   identified   in
Sec.~\ref{sec:method}.

\begin{figure}
  \includegraphics[width=\columnwidth]{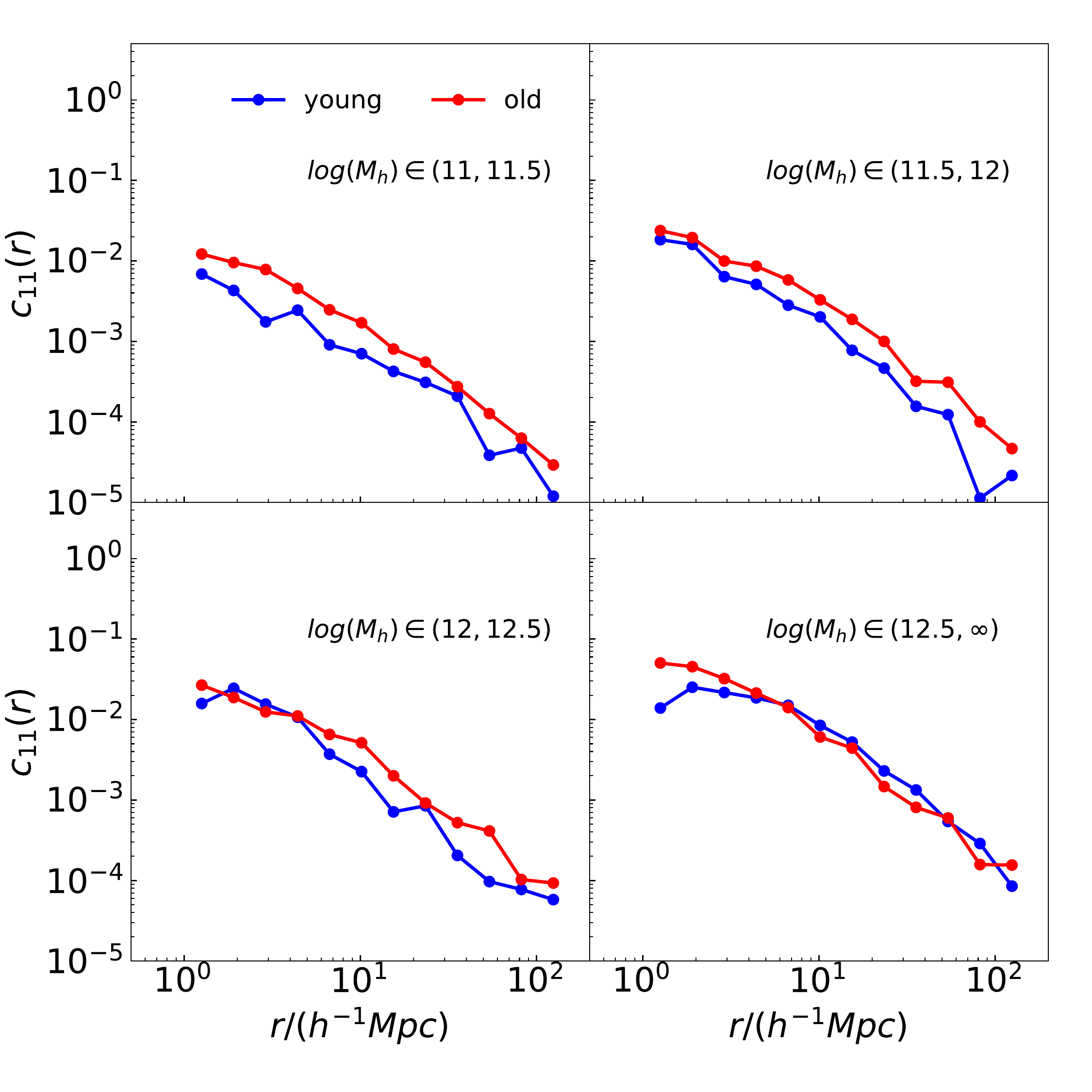}
  \caption{Unweighted halo ellipticity correlation function $\coneone$ ($q=0$) at $z=0$ for old and young halos in a few mass bins. The difference between their correlation is decreasing with halo mass, consistent with the age dependence of halo bias as function as halo mass.}
    \label{fig:ftime_diff}
  \end{figure}

In Fig.~\ref{fig:ftime_diff}, we plot the unweighted $\coneone$ signal
for  old and  young halos  in a  few  mass bins.  It is  seen that  in
low-mass bins  ($\lg M<12.5$), old  halos have stronger  alignment than
young halos and  at intermediate scales $\sim 10Mpc/h$,  the IA of old
halos is around  2 times of the young halos.  For massive halos (lower
right panel)  the difference  between the two  samples is  small. This
trend   can    be  quantitatively explained    by   our   tidal    alignment   model
(Eq.~\ref{eqn:theory}) with inclusion of halo bias. These difference of
halo  IA  in old  and  young  halos are  in  good  agreement with  the
dependence     of     halo    bias     on     formation  time as  found     by
\cite{2004MNRAS.355..819G}.

Previously, to  explain the mass dependence of halo IA, \cite{2005MNRAS.360..203S} argued that massive halos are stronger aligned as they are  formed later, so they have less time to virialize  and still keep the  memory of the tidal  field that they collapsed within. Our results above show  that halo IA is dependent on both halo mass and the formation time, and it can  be understood using our linear model in terms of halo bias with its dependence on mass and formation time separately.    

Many  studies have  shown  that  halo spins  are  correlated with  the
large-scale     environment     (e.g.,     \cite{2007MNRAS.375..489H},
\cite{2015ApJ...813....6K} and  references therein).  There  is also a
strong possibility that  halos IA will also depend  on the large-scale
environment.   In  this part  we  compare  the  halo IA  in  different
environments,  mainly in cluster  and filament.  We select question halos with
mass  $>10^{12}\hsolar$,  and label  those  in  cluster and  filament
  environment   as  pure   $Q_\text{clu}$   and  pure   $Q_\text{fil}$ sample 
  respectively.  As  halo  mass   is  strongly  correlated  with  its
  environment (e.g.,\cite{2007MNRAS.375..489H})  that halos in cluster
  environment  are  usually larger,  therefore,  direct comparison  of
  $\coneone$ between  these two  samples will suffer  from significant
  mass-dependence effect as seen in previous section.  So we construct
  two     control    samples,     $Q_\text{clu}^\text{control}$    and
  $Q_\text{fil}^\text{control}$  respectively, ensuring they  have the
  same halo  mass distribution  as the pure  cluster/filament samples,
  but regardless of their environment.

\begin{figure}
\includegraphics[width=\columnwidth]{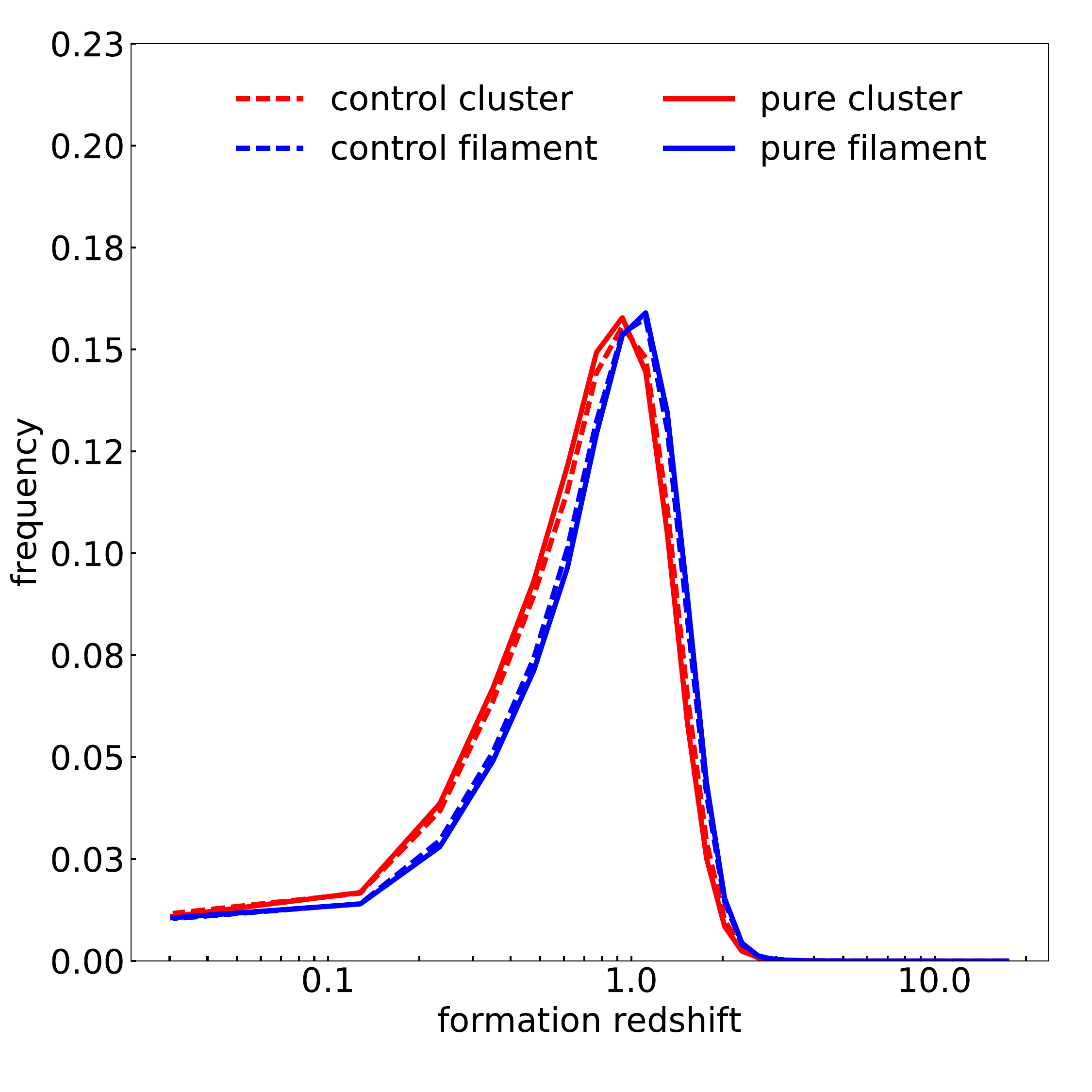}
\caption{Formation time distribution for question (solid) and control (dashed) samples of halos in cluster and filament environment.}
  \label{fig:ftime_LSS}
\end{figure}

After  eliminating  the halo  mass  dependence,  to  see if  there  is
additional difference  in the formation time between  the question and
control  samples,   we  plot  their  formation   time  distributions  in
Fig.~\ref{fig:ftime_LSS}. By comparing the solid and dashed lines, it  is  seen   that  under  the  same  mass
distribution the formation time  is identical between the question and
control sample.  Previous  work  (e.g.,   \cite{2007MNRAS.375..489H})  have  found  for
low-mass halo  ($<10^{12}\hsolar$), the formation  time has dependence on the environment, in which  cluster halos form earlier than those in
filaments and voids,  but for massive halos there  is no environmental
dependence. As  we are looking at halos  larger than $10^{12}\hsolar$,
our results agree well with their results.
The formation time  of halos in cluster environment is
slightly lower than the halo in filaments, this is because the average
halo  mass in  cluster  is larger,  so  the formation  time is  lower.

\begin{figure}[t]
\includegraphics[width=\columnwidth]{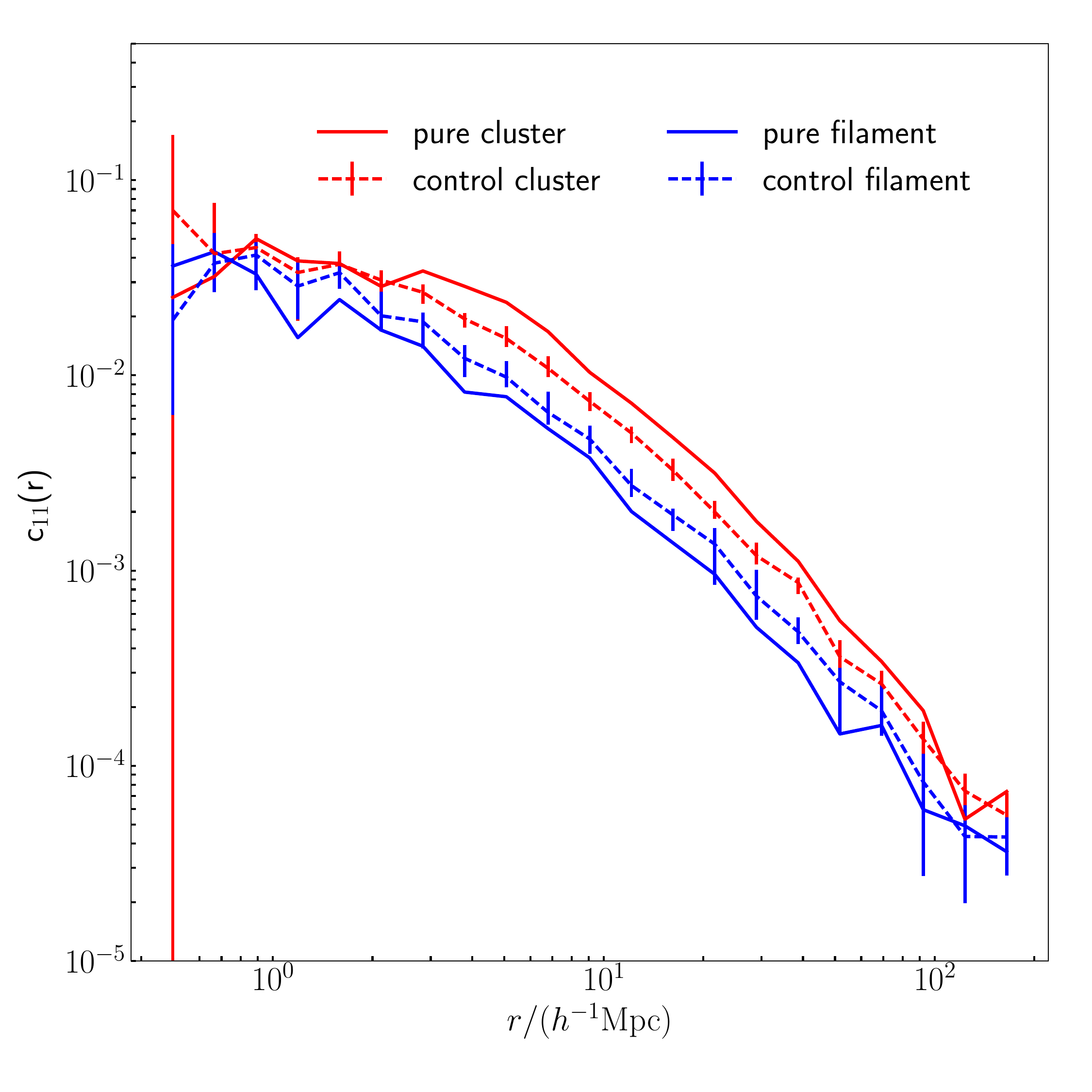}
  \caption{Ellipticity correlations (unweighted) at $z=0$ for halos in cluster and filament environment. The results for question samples are shown in solid lines and control samples in dashed lines. The control samples have the same halo mass distribution as the corresponding question samples. The error bars are from 10 realizations of the control samples.}
  \label{fig:c11_LSS}
\end{figure}

In Fig.\ref{fig:c11_LSS}, we plot $\coneone$ for the pure cluster/filament samples in solid lines and the control samples in dashed lines. The errorbars are taken from 10 realizations of the control samples. We find that under the same mass distribution, $\coneone$ from halos in cluster environment (red solid line) is much stronger than the control sample which has a mixture of different  environments (red dashed line). On the other hand, the $\coneone$ for halos in filament (blue solid) is much weaker than that in the control samples (blue dashed). The difference between solid lines includes both mass and environment dependence. The fact that $\coneone$ for $Q_\text{clu}^\text{control}$ is higher than that of $Q_\text{fil}^\text{control}$ is mainly due to the mass dependence.

\begin{figure}[t]
    \centering
    \includegraphics[width=\columnwidth]{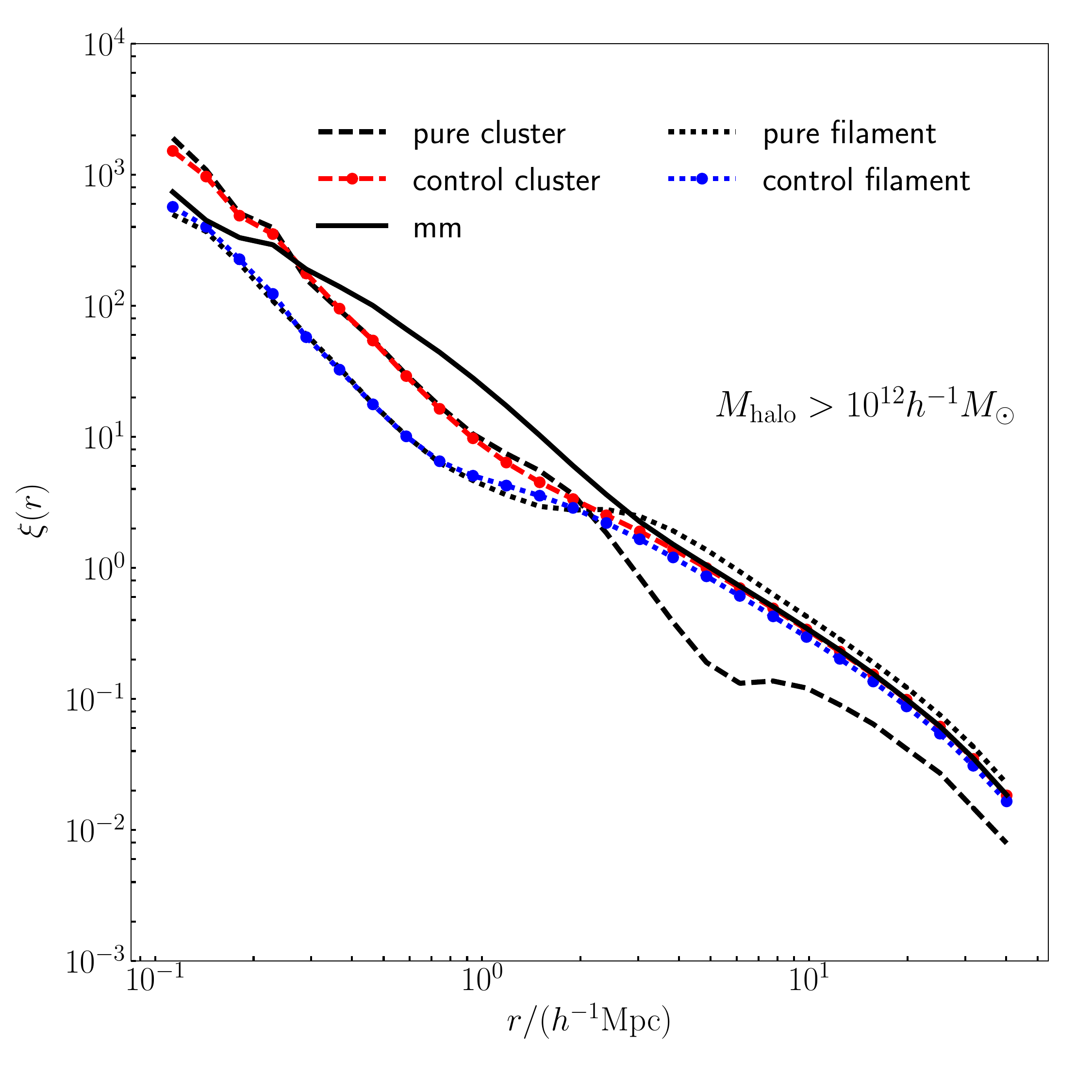}
    \caption{The  cross-correlations in different cosmic environment. The black dashed and dotted lines are the halo-mass cross-correlation for halos in cluster and filament, and red/blue lines are for the corresponding control samples. The black solid line is the auto-cross correlation of the background mass distribution. Clear cosmic environment dependence on large-scale is seen that cluster halos have lower clustering (bias) than the background mass distribution, and filament halos have higher bias.}
    \label{fig:bias_LSS}
\end{figure}

In order to understand the origin of environment dependence of halo IA
and to see  if it can be  explained by halo bias as  introduced in our
theoretical  model in  Sec.~\ref{sec:method}, here  we  investigate the  halo bias  in different environment.  We calculate the cross-correlation function $\xi_{hm}$ between the halos and the dark matter particles. For references we also calculate the auto-correlation function $\xi_{mm}$ of the background particles.

In Fig.~\ref{fig:bias_LSS}, we  show the cross-correlations  of halos in different environment with the background particles. Firstly  we note that, below the scale $\sim 2\hmpc$,  the  halo-matter cross-correlation  function mainly describes the    clustering   of   dark    matter   particles   inside
halos. Therefore, this part of the correlation function mainly depends
on the average halo mass in the studied sample. Since the question and
control samples  have the same  mass distribution, the  clusterings on
small scales are very similar between them.

On  scales beyond  $2\hmpc$, the  cross-correlation function
then reflects the level  of clustering between particles in different
halos. It  is found  that halos in  cluster environment  (black dashed
line)  is less  clustered than  halos in  filament  environment (black
dotted  line). The  control  samples have  similar  clustering as  the
corresponding question  samples on small  scales, but on  large scales
they are similar  to the background dark matter  particles, and the control
cluster sample (red dashed line) has slightly higher correlations than
the control  filament sample  (blue dashed line)  due to  their higher
average halo mass.

\begin{figure*}[t]
  \begin{center}
  \includegraphics[width=\hsize]{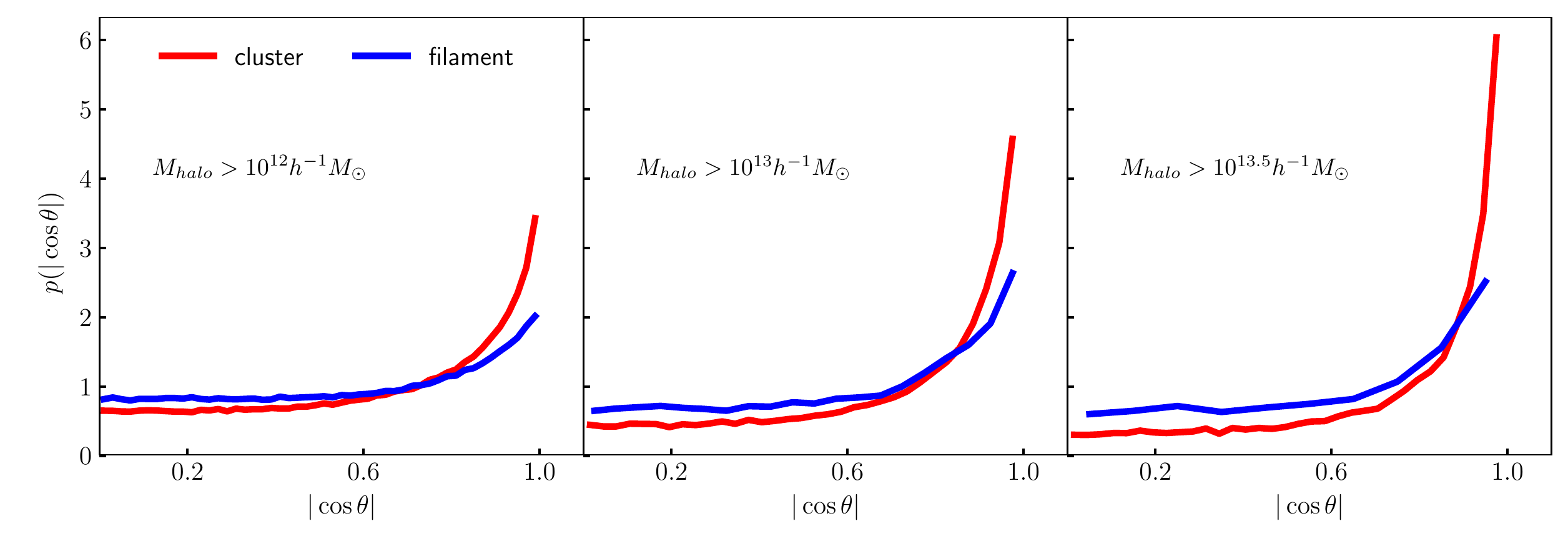}
  \end{center}
    \caption{The probability density function of the halo-tidal alignment in cluster (red) and filament (blue) environment in mass bins $M_{\text{halo}}> 10^{12},10^{13},10^{13.5} h^{-1}M_\odot$ respectively.}
    \label{fig:tidal_align}
\end{figure*}
The higher clustering on large  scales for halos in filament than halos 
in cluster is unexpected. We note  that it is not due to the formation
time difference between the two sample,  as Fig.~\ref{fig:ftime_LSS} shows
that their formation time distributions are very similar. Interestingly,
in a  recent paper by \cite{2016arXiv161004231B} from  the ZOMG project,
the author found that stalled  (filament) halos has a larger halo bias
$b_h>1$  than the  accreting  (cluster)  halos with  $b_h  < 1$.   The
environment  dependence is  also  recently confirmed  by  \cite{2017arXiv170402451Y} who  study the halo bias  in much more detail  using the ELUCID
simulations, and they also find  that cluster halos are less clustered
on  larger  scale. More recently, \cite{2017arXiv170609906P} has found some very interesting results. Although they used a different method to define the large-scale environment (the anisotropy factor $\alpha_R$), they found that halo bias evolves smoothly with $\alpha_R$, and in particular, halos in ‘isotropic’ (node) environment have a negative bias while those in ‘anisotropic’ (filament) environment have a positive bias (see their Fig.12). This is qualitatively in agreement with our results in Fig.~\ref{fig:bias_LSS}. In  general,  our results  agree  with the  recent
findings.

The fact that halo bias is higher in filament and lower in cluster
is  opposite   to  the  trend   of  environment  dependence   seen  in
Fig.~\ref{fig:c11_LSS}, and  therefore we  cannot use  the halo bias term in our theoretical model to explain the environment dependence in
halo  IA.  These  two  opposite  trends  for  large-scale  environment
dependence also suggests that additional factors other than halo bias must be
taken into account to understand this difference. In particular, these
factors need  to produce a significant difference in halo IA  between cluster and
filament  to overcome any  halo bias  effect and  result in  the final
$\coneone$ that we see in Fig.~\ref{fig:c11_LSS}.

One of such  potential factors is the level  of alignment between halo orientation 
and     the     large-scale     environment. The basic assumption in the linear alignment model (Hirata \& Seljak 2004) is that halo shape is perfectly aligned with the tidal field.    As     reported     in
\cite{2015ApJ...798...17Z},  and \cite{2015ApJ...813....6K}, although the halo
shape  is strongly  correlated with  the  local tidal  field, there is still a mis-alignment with dependence on halo mass and environment. Here  we
calculate the alignment  between the major axis ($v_{1}$)  of the halo
shape and  the direction  of the tidal  field, i.e., $e_{3}$  from the
Hessian  matrix,  which  is  the  direction of  the  slowest  collapse
direction  (for detail,  see Kang  \& Wang  2015). We  calculate their
alignment as, 
\begin{align}
{\bf e}_3 \cdot {\bf v}_1 = \cos\theta    
\end{align}
where larger 
$|\cos\theta|$ means that the halo major axis is better aligned with the tidal field.

In  Fig.~\ref{fig:tidal_align},   we  plot  the  probability  distribution   
of   $|\cos\theta|$   for   halos   in   three   mass   bins
$M_{\text{halo}}>  10^{12},10^{13},10^{13.5}  h^{-1}M_\odot$  and in  both
filament and cluster  environment. It is found that  halos in clusters
have  a  stronger  alignment  with  local  tidal  field  in  all  mass
ranges. This result, given that large scale tidal field correlation underlies
the  large  scale  IA  (\cite{2001MNRAS.323..713C}) ,  provides  a
plausible reason of why the halo  IA being stronger in clusters and weaker
in filaments. We  believe this effect should be  taken into account to
fully   explain  the  environment   dependence  of   halo's  intrinsic
alignment. In this paper we do not study this effect in details.
  \begin{figure*}
  \begin{center}
  \epsfig{figure=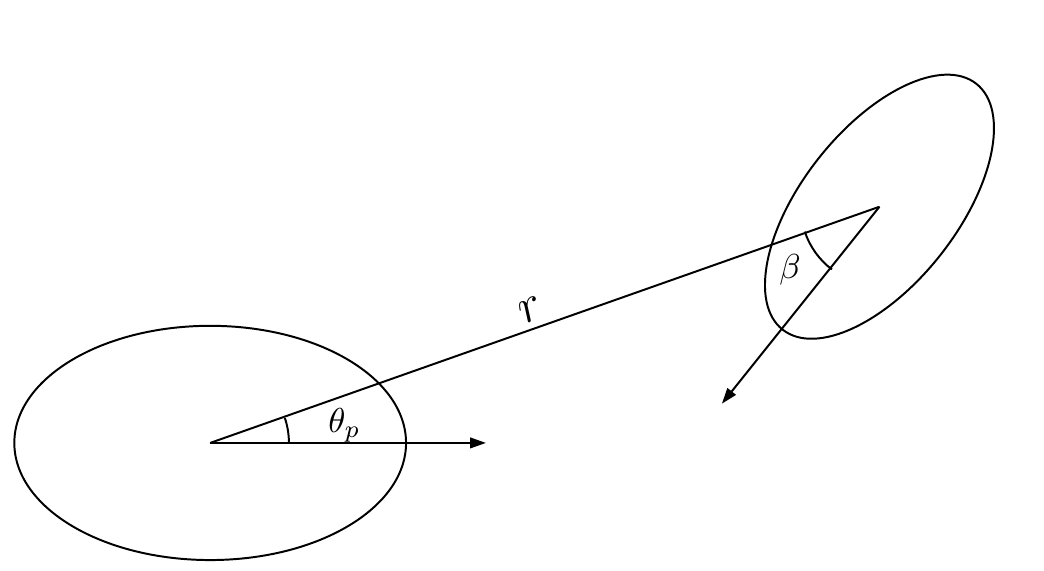,width=0.47\hsize}
  \epsfig{figure=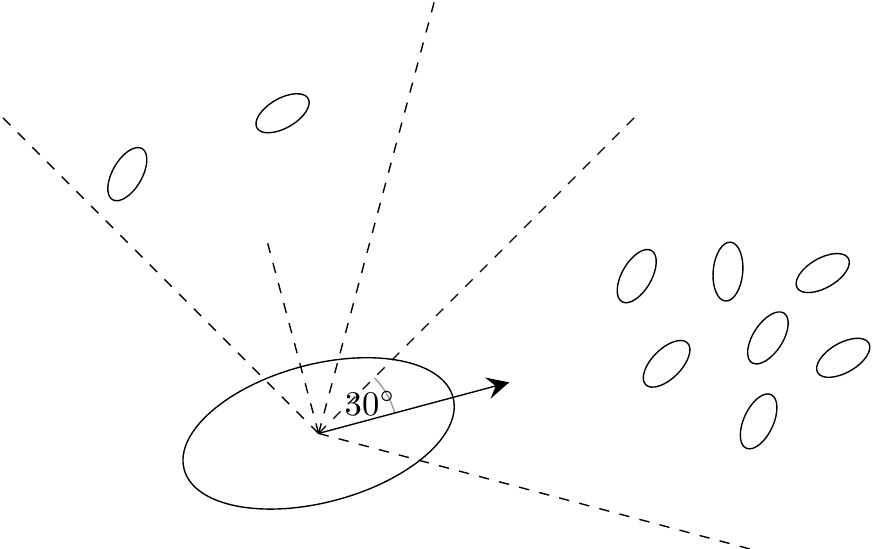,width=0.47\hsize}
  \caption{Left: Illustration of the pointing angle $\theta_p$ and $\beta$; Right: Exaggerated illustration of the alignment correlation function.}
  \label{fig:illustration1}
  \end{center}
  \end{figure*}

\section{The Alignment Correlation Function}
\label{sec:ACF}
\begin{figure}[t]
    \includegraphics[width=\columnwidth]{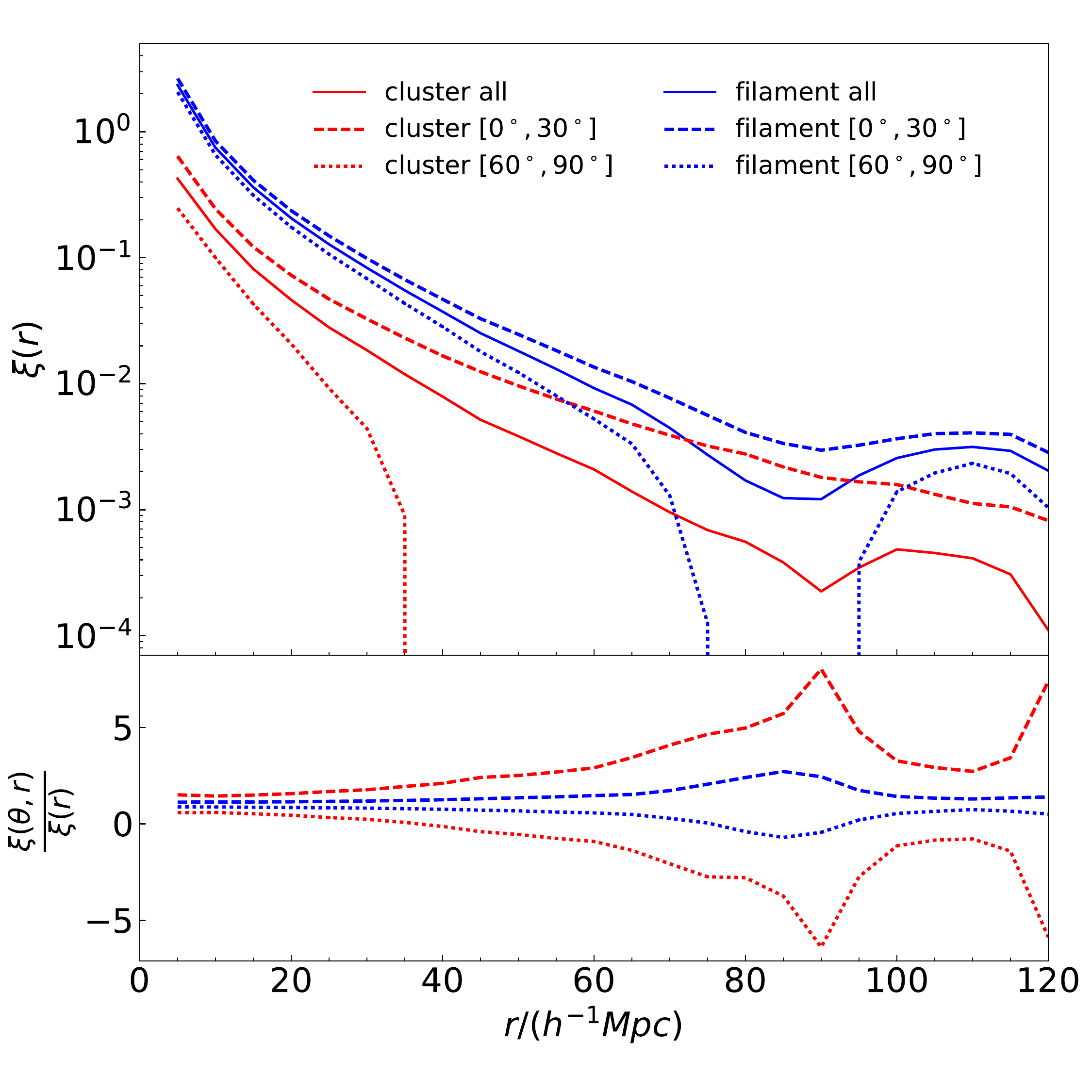}
  \caption{The halo alignment correlation functions. Upper panel: halo ACFs along different alignment angles. Red lines are for halos in cluster environment and blue for filament environment. Dashed lines are for correlations along the major axes of the halos ($0^\circ < \theta_p < 30^\circ$) and dotted lines are for correlations along the minor axes of the halos ($60^\circ < \theta_p < 90^\circ$). Black lines are for halos with no limit on the alignment angle. Lower panel: the ratio between halo ACF along different angles to the halo correlations with no limit on the alignment angle.}
  \label{fig:ACF_redist}
\end{figure}

In  Sec.\ref{sec:c11}, we  have shown  the dependence  of halo  IA and
comparison   between   simulation   results   with  theoretical   model
predictions. There the halo IA is  defined in a way similar to the one
often used  in weak lensing analyses. In fact the halo alignment has been measured in 
another way, called alignment correlation function (ACF), in which the
halo clustering is a function of the direction along one halo axis (e.g.,  \cite{2009RAA.....9...41F}). Although the
ACF  and halo IA can be
connected (especially GI term \cite{2011JCAP...05..010B}), the ACF provides more direct insight on
how  halos/galaxies are  correlated  along  their major or minor  axes.   An  illustration   of   the  ACF   statistic   is  shown   in
Fig.~\ref{fig:illustration1} where  $\theta_{p}$ is the  angle between
the major axis of a halo and the separation vector to another halo.

Given a sample of halos in question (sample Q) which is a subset of the reference sample G, the ACF $\xi(\theta_p,r)$ is defined as (\cite{2009RAA.....9...41F}),
\begin{align}
\xi(\theta_p,r) = \frac{N_R}{N_G}\frac{QG(\theta_p,r)}{QR(\theta_p,r)} - 1 ,    
\end{align}
where R is a random sample. 

The  halo  and  galaxy  ACFs  have  been measured  in  a  few  studies
(\cite{2009RAA.....9...41F}, \cite{2012MNRAS.423..856S}, \cite{2013ApJ...770L..12L}, 
\cite{2016MNRAS.461.2702C}, \cite{2015MNRAS.454.2736C}). \cite{2013ApJ...770L..12L}
showed the direct measurement of the ACF using CMASS galaxy catalogue,
and in  the meantime  calculated the ACF  for dark matter  halos using
cosmological simulation.  They found  evidence that galaxies and halos
are more clustered than the  average within a spanned region $\theta_p
= (0^\circ, 30^\circ)$ around the major axis and less clustered within
$\theta_p =  (60^\circ, 90^\circ)$ at almost  all scales. Furthermore,
they  found the  correlation  can be  seen  up to  very large  scales at around $70 \hmpc$. The results of \cite{2013ApJ...770L..12L} motivate  us to  study halo  ACFs with  their  dependence on
environment  and investigate  the origin  of the  correlation  on very
large scales.

Here we present an investigation on the environment dependence of halo  ACF. In order to look  for signals on very  large scales, we
use  the  full   halo  sample  in  the  C4   simulation  with  box-size of
$1h^{-1}{\rm Gpc}$  at $z=0.6$ (to match the \cite{2013ApJ...770L..12L} data from observations)
as the reference sample G ($N_G  = 12,640,839$), and we use the set of
halos  with mass  $M_h \ge  10^{12}\hsolar$  as sample  Q. The  random
sample R is generated with $N_R/N_G = 10$. We divide halos in sample Q
into two samples, $Q_{clu}$  and $Q_{fil}$, based on their environment
identified    as    cluster    and    filament. In  the upper panel  of Fig.\ref{fig:ACF_redist}
we  plot the  ACFs for  these samples.

Fig.~\ref{fig:ACF_redist} shows a few interesting features. First, it is found that without limitation on the alignment angle $\theta_p$, the halo ACF in cluster (red solid line) is lower than halos in filament (blue solid line). This is consistent with the results in Fig.~\ref{fig:bias_LSS} although there it plots the cross correlation between halo and dark matter and here we show the halo-halo cross-correlation. The halo ACF also displays a peak at around $110 \hmpc$, which is exactly the signal from the baryon acoustic oscillations (BAO). Secondly, it is found that in agreement with \cite{2013ApJ...770L..12L} the halo ACF with $0^\circ < \theta_p < 30^\circ$ (dashed lines) is higher in both filament and cluster environment than the average, indicating that halo clustering is enhanced along their major axes. Accordingly, the clustering is decreased with $60^\circ < \theta_p < 90^\circ$.  An illustration of this effect is seen in the right panel of Fig.~\ref{fig:illustration1}. 

In the lower panel of Fig.~\ref{fig:ACF_redist} we show the ratio between halo ACFs along different alignment angles with the average ACFs in the cluster and filament environment, respectively. The dashed lines are for halos clustering along the major axes ($0^\circ < \theta < 30^\circ$) and the dotted lines are for results along the minor axes of the halos ($60^\circ < \theta_p < 90^\circ$). It is seen that although the halo ACF is lower in cluster, but the correlation along the major axis is significantly enhanced in cluster than in filament. It is also interestingly found that the clustering along the minor axes will become negative at some scales with dependence on halo environment. For example, for halos in cluster the ACF is negative at $r>35 \hmpc$ and reaches the lowest dip at around $90 \hmpc$ after which it increases to the peak at around $110 \hmpc$. For halos in filament, the clustering along minor axes becomes negative at around $70 \hmpc$ and reach a dip at around $90 \hmpc$. These results indicate that halo in clustering environment has a strong dependence on the alignment angle. 

In addition to the halo ACF, we also calculate the alignment signal $\ctt$ over correlated pairs, which is defined in \cite{2009RAA.....9...41F} as 
\begin{equation}
\langle\cos(2\theta_p)\rangle_\text{cor}(r) = \frac{\int_0^{\pi/2} \cos(2\theta_p) \xi(\theta_p,r) {\rm d}\theta_p} {\int_0^{\pi/2} \xi(\theta_p,r) {\rm d}\theta_p} ,
\label{eq:cos2theta_def}
\end{equation}
and is estimated by
\begin{equation}
\langle\cos(2\theta_p)\rangle_\text{cor}(r) = \frac{QG_{\theta_p}(r)}{QG(r)-(N_G/N_R)\times QR(r)} ,
\label{eq:estimator}
\end{equation}

\begin{figure*}
  \begin{center}
    \epsfig{figure=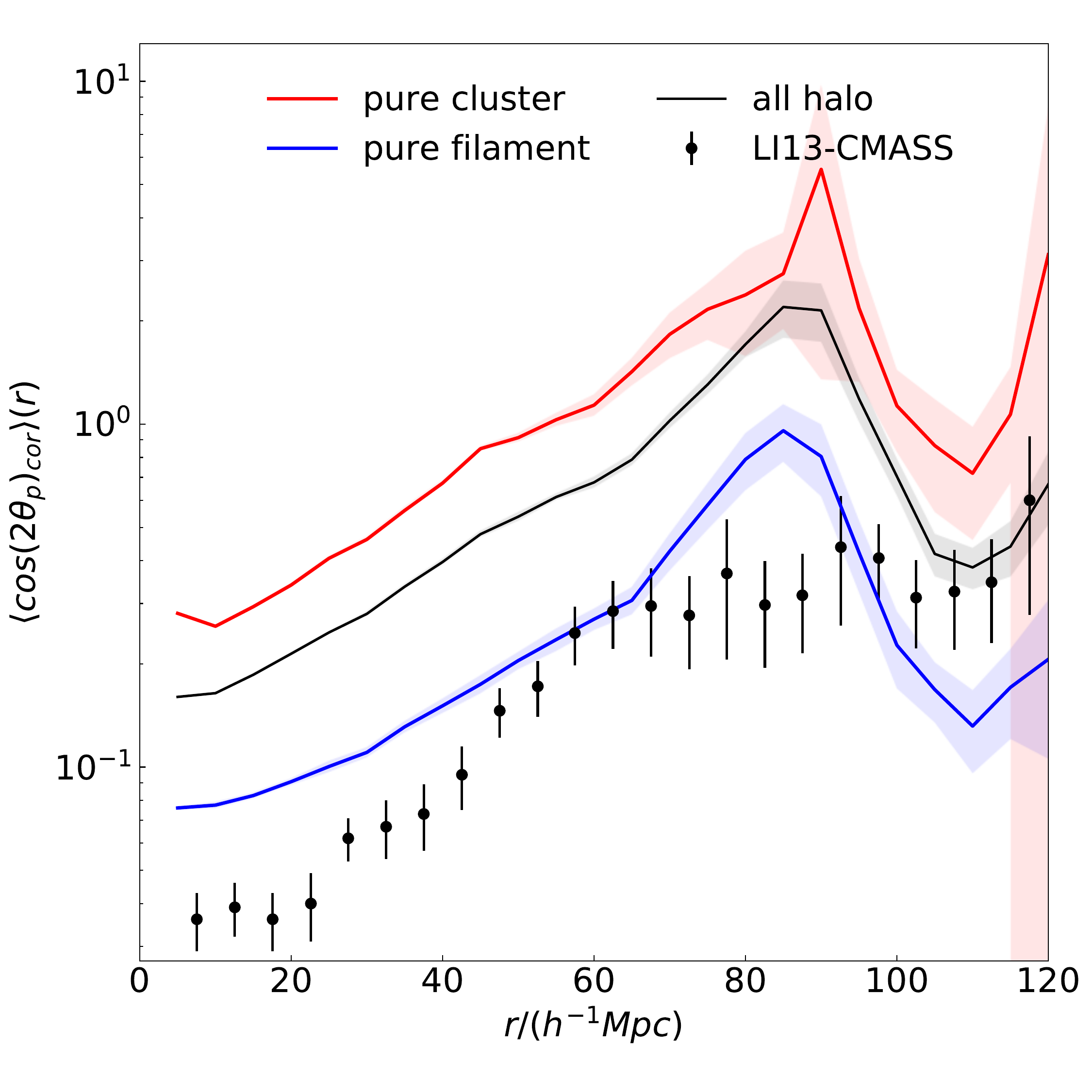,width=0.47\hsize}
    \epsfig{figure=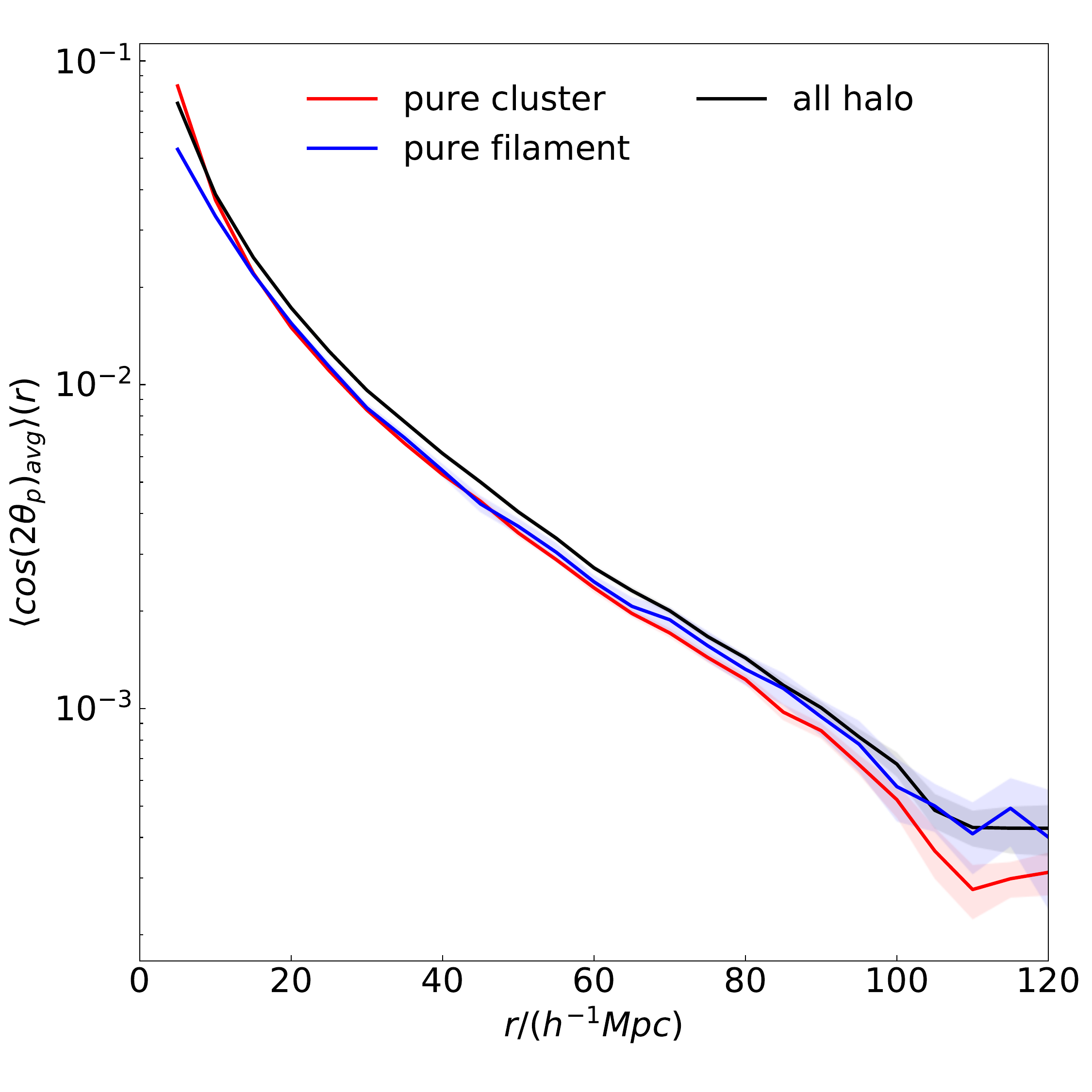,width=0.47\hsize}
  \end{center}
  \caption{The correlated and average halo alignment angles. Left: The correlated alignment angle, $\cos(2\theta_p)_{\rm cor}$, for correlated halo pairs in filament (blue), cluster (red) and all environment (black). The data points are from measurement in \cite{2013ApJ...770L..12L} using CMASS galaxies. Right panel: the average alignment angle,$\cos(2\theta_p)$, for all halo pairs in different environment. }
  \label{fig:ACF_cos2theta}
\end{figure*}

In the left panel of Fig.~\ref{fig:ACF_cos2theta}, we show the calculated $\corr{\ctt}_{\rm cor}$ for above samples together with CMASS measurement from Li et al. (2013). It is seen that for all halo sample (black line), the ACF is larger than observed galaxy ACF (data points). This is not surprised, as Li et al. (2013) have obtained similar results and they also showed that a mis-alignment between galaxy orientation and halo orientation has to be included so as to dilute the halo ACF to agree with the data. As here we are focusing on the environmental dependence, we do not show such a test. 

What is interesting in the left panel of Fig.~\ref{fig:ACF_cos2theta} is that the ACF is stronger in cluster environment than the filament environment and both have oscillations with a peak around $90 \hmpc$. This environment dependence seems to be inconsistent with the above results that halo ACF is stronger in filament environment. In order to understand the physical origin of the oscillation of the $\corr{\ctt}_{\rm cor}$ on large scales and the environment dependence in more detail, we take a further look at the $\cos(2\theta_p)_{\rm cor}$ statistic and find  that the estimator in Eq.~\ref{eq:cos2theta_def} can be decomposed into two parts. By recognizing the two-point cross-correlation function as $\xi = \frac{N_R}{N_G}\frac{QG}{QR} - 1$, the estimator of $\corr{\ctt}_{\rm cor}$ statistic is equivalent to 
\begin{align}
\corr{\cos(2\theta_p)}_\text{cor} &= \frac{QG_{\theta_p}}{QG} \frac{\xi+1}{\xi} \nonumber \\
 &= \corr{\cos(2\theta_p)}_\text{avg} (1+\frac{1}{\xi})
\label{eq:cos2theta}    
\end{align}
where $\corr{\ctt}_\text{avg}$ is the mean value of $\ctt$ over all halo pairs instead. 

The above equation shows that the correlated halo alignment $\cos(2\theta_p)_{\rm cor}$ is determined by two components, one is the halo correlation and the other is the average halo alignment angle in all pairs. In region where the halo correlation $\xi$ is very small ($\xi << 1$), the correlated alignment signal $\cos(2\theta_p)_{\rm cor}$ is the average alignment $\cos(2\theta_p)$ divided by the halo correlation. We have shown the one component, the clustering for halos $\xi(r)$, in filament and cluster in Fig.~\ref{fig:ACF_redist} and here we plot the other component, average halo alignment angle, for halos in different environment in the right panel of Fig.~\ref{fig:ACF_cos2theta}. 

The right panel shows that the average alignment angles of halos are very similar between filament and cluster environment, and it decreases with halo separation. As the clustering of halo in cluster environment is much lower than that of filament (red and blue solid lines in Fig.~\ref{fig:ACF_redist}) and $cos(2\theta_p)_{\rm cor}$ is proportional to $1/\xi(r)$, so it leads to a higher correlated halo alignment angle in clusters. The oscillation in the clustering at $90 \hmpc < r < 120 \hmpc$ due to the BAO effect also explains the peak and dip of halo correlation alignment seen in the left panel of Fig.~\ref{fig:ACF_cos2theta}. It is also interesting to note that the correlated halo alignment $\cos(2\theta_p)_{\rm cor}$ in cluster is higher than $1$ at $r \sim 90 \hmpc$. This seems to be impossible from the definition in Eq.~\ref{eq:cos2theta_def} where it implies that the statistic should range between -1 and 1. However, it is true only when the halo ACF ($\xi(\theta_p,r)$) in Eq.~\ref{eq:cos2theta_def} is always positive. We have seen from Fig.~\ref{fig:ACF_redist} that the halo ACF is negative for $60^\circ < \theta_p < 90^\circ$ at some distances, so the product of two negative components, $\xi(\theta_p,r)$ and $\cos(2\theta_p)$, leads to a positive value and results in a final $\cos(2\theta_p)_{\rm cor}$ being larger than 1 at some distances.

Finally, we note the seeming inconsistency between the environmental dependence of $\coneone$ (higher in cluster) and the ACF (higher in filament). The ACF is the clustering along the halo major axis, so ACF is mainly dominated by halo clustering. In Fig.~\ref{fig:bias_LSS} we have shown that halo clustering is higher in filament, so the ACF is also higher in filament than in cluster. On the other hand, $\coneone$ (unweighted by halo ellipticity) measures the average of the product of ``$\cos2\theta_p$'' between halo pairs, and it can be written as,

\begin{align}
c_{11}  &= <\cos2\theta \cos2\beta> \\
        &= \int_0^{\pi/2}\int_0^{\pi/2} \cos2\theta_p \cos2\beta \frac{QQ(\theta_p,\beta,r)}{QQ(r)} {\rm d}\theta_p{\rm d}\beta \\
        &= \int_0^{\pi/2}\int_0^{\pi/2} \cos2\theta_p \cos2\beta \frac{1+\xi(\theta_p,\beta,r)}{\xi(r)} {\rm d}\theta_p{\rm d}\beta,
        \label{eqn:theta_beta}
\end{align}
where the definition of $\theta_p$, $\beta$ can be found from the halo configuration on the left of Fig.~\ref{fig:illustration1}. It is seen that $\coneone$ arises from two terms. The term $(1+\xi(\theta_p,\beta,r))/(1+\xi(r))$ is the enhancement of clustering along halo major axis. We have shown in Fig.~\ref{fig:tidal_align} and lower panel of Fig.~\ref{fig:ACF_redist} that in cluster environment, halo major axis is better aligned with the tidal field and the clustering is significantly enhanced along the halo major axis than those in filament. We also found that the term $cos2\theta \cos2\beta$ is higher for halo pairs in clusters than those in filament. Therefore, the environment dependence of $\coneone$ and ACF enhancement are consistent with each other, as both can be explained by the alignment between halo major axis and tidal field.

\section{Summary and Conclusion}
\label{sec:Conclusion}
Intrinsic alignment of galaxies is a result of complicated physics of galaxy formation and the tidal interaction of dark matter halo with the large-scale structure. Galaxy intrinsic alignment has demonstrated its importance as a challenge to weak lensing in the era of precision cosmology. Understanding the dependence of galaxy intrinsic alignment on galaxy properties is a key step to model its effect on the cosmic shear and other useful statistics in the weak lensing survey. As a first step it is important to understand the halo intrinsic alignment with its dependence on mass, formation time and large-scale environment. In this paper we make use of two N-body simulations to investigate these dependences and the main results are summarized as followings:
\begin{enumerate}
	\item Intrinsic alignment of dark matter halo has long been found to be higher for massive halos in N-body simulations (\cite{2002MNRAS.335L..89J}). However, the halo alignment and its mass dependence are often neglected in the linear model for galaxy alignment and it is not clear whether the linear tidal model (e.g., \cite{2004PhRvD..70f3526H}) can describe the halo intrinsic alignment and its mass dependence.  By modifying the linear alignment to its 3D form and including the halo bias term, our slightly modified model is able to predict the mass dependence for halo intrinsic alignment in simulations. The redshift dependence can also be closely followed using the nonlinear power spectrum at different redshift.
	\item We find that for given mass, older halos have stronger alignment than younger halos. The trend in halo formation time can be well captured by our modified linear alignment model and the difference in formation is consistent with that of halo bias found by \cite{2004MNRAS.355..819G}. Our results confirm that halo formation time is an independent factor in determining the intrinsic alignment. 
	\item We have measured the halo alignment correlation function from the C4 simulation. In agreement with previous results (e.g., \cite{2009RAA.....9...41F}; \cite{2013ApJ...770L..12L}), we find that the halo clustering is significantly enhanced along the halo major axes and is less clustered along the minor axes. 
  \item For halos in different large-scale environments, the $\coneone$ is found to be higher in cluster and lower in filament. A further look into the dark matter-halo cross-correlation function reveals an opposite trend that halo bias is stronger in filament than in cluster. Such a bias with dependence on environment is consistent with recent findings (e.g., \cite{2017arXiv170609906P}, \cite{2017arXiv170402451Y}). This opposite trend in halo bias can explains the environment dependence in ACF, but not the $\coneone$. We find that the enhancement of ACF in cluster is much more stronger than in filament, which can explain the environmental dependence in $\coneone$. The enhancement of ACF along the halo major axes is due to the alignment between halo shape and the tidal field and this effect is stronger in cluster.  Our results suggest that, beside the halo bias, the linear tidal model for halo alignment should also include a factor to describe the correspondence between halo shape and the local tidal field, as well as its environmental dependence.
  \item Furthermore, we have found that the alignment angle between correlated halo pairs has a peak at around $90\hmpc$ and a dip at around $110\hmpc$. We pointed out that such an oscillation is caused by the baryon acoustic oscillation.
\end{enumerate}

Finally, we note that the study of halo IA still needs to be extended in future work to model galaxy IA in order to shed more light on weak lensing observations. In the language of halo model, this paper mainly focused on two-halo term and therefore does not include any recipe for the one-halo alignment for satellite galaxy. For a detailed discussion of the effect of satellite alignment on observed lensing signal, we refer readers to Wei et al. (2017, in prep). On the other hand, the formation time dependence on halo IA can be an important feature to look for in galaxy surveys. Observations have found the ACF signal using CMASS galaxies, and the environment dependence on ACF can be tested using galaxy catalogues with LSS classification, e.g., GAMA. In general, the above dependencies in alignment statistics can be extended to galaxy IA statistics, which would be useful for future development in galaxy IA both theoretically and observationally.

\acknowledgments
{\bf Acknowledgement:} We would like to thank the anonymous referee for the constructive comments which significantly contributed to improving the quality of the publication. QX thanks Jonathan Blazek, Catherine Heymans and John Peacock for helpful discussions and hospitality at the Royal Observatory of Edinburgh.  QX also thanks Zhe Chu for carefully reading of the manuscript. The work is supported by the NSFC (No. 11333008, 11233005, 11621303, 11522324, 11421303), 973 program (2015CB857003, 2015CB857002, 2013CB834900), the NSF of Jiangsu Province (No. BK20140050). The ELUCID simulation is run using at the Center for High Performance Computing, Shanghai Jiao Tong University.


\bibliographystyle{apj}

\bibliography{ms}

\end{document}